\documentclass[fleqn,usenatbib]{mnras}

\usepackage{mathptmx}

\usepackage[T1]{fontenc}
\usepackage{ae,aecompl}

\usepackage{graphicx}	
\usepackage{amsmath}	
\usepackage{amssymb}	
\usepackage[usenames,dvipsnames]{color} 
\usepackage{mathrsfs}
\usepackage{verbatim}
\usepackage[normalem]{ulem}

\def\araa{ARA\&A}%
\def\apj{ApJ}%
\def\apjl{ApJ}%
\def\apjs{ApJS}%
\def\ao{Appl.~Opt.}%
\def\apss{Ap\&SS}%
\def\aap{A\&A}%
\def\mnras{MNRAS}%
\newcommand{\Msun}{\mbox{M$_\odot$}} 	

\newcommand{\Mhef}{\mbox{$M_\mathrm{He,0}$}}	
\newcommand{\Mhe}{\mbox{$M_\mathrm{He}$}}	

\newcommand{\rcc}{\mbox{$r_\mathrm{cc}$}}	 

\newcommand{\gradT}{\mbox{$\nabla_{\mathrm{T}}$}}	
\newcommand{\gradr}{\mbox{$\nabla_{\mathrm{r}}$}}
\newcommand{\grada}{\mbox{$\nabla_{\mathrm{a}}$}}

\newcommand{\feh}{\mbox{[Fe/H]}}		

\newcommand{\amlt}{\mbox{$\alpha_{\mbox{mlt}}$}}	
\newcommand{\aovhe}{\mbox{$\alpha_{\mbox{ovHe}}$}} 

\newcommand{\dnu}{\mbox{$\Delta\nu$}}		
\newcommand{\numax}{\mbox{$\nu_\mathrm{max}$}}   
\newcommand{\dpg}{\mbox{$\Delta\Pi_\mathrm{1}$}} 
\newcommand{\dpgmax}{\mbox{$\Delta\Pi_\mathrm{1,max}$}} 
\newcommand{\dpgmin}{\mbox{$\Delta\Pi_\mathrm{1,min}$}} 

\newcommand{\coreaction}{\mbox{$^\mathrm{12}$C($\alpha$,$\gamma$)$^\mathrm{16}$O}}
\newcommand{\trialfa}{\mbox{triple-$\alpha$}}

\definecolor{dgreen}{RGB}{0,127,0}
\definecolor{gray}{gray}{0.5}

\newcommand{\schw}{Schwarzschild}	
\newcommand{\kepler}{{\it Kepler}}
\newcommand{\brunt}{Brunt-V\"ais\"al\"a}
\newcommand{\etal}{\mbox{\rm et~al.}}		

\newcommand{\beq}{\begin{equation}}		
\newcommand{\eeq}{\end{equation}}		
\newcommand{\beqa}{\begin{eqnarray}}	
\newcommand{\eeqa}{\end{eqnarray}}	
\newcommand{\bitem}{\begin{itemize}}
\newcommand{\eitem}{\end{itemize}}	
\newcommand{\benum}{\begin{enumerate}}	
\newcommand{\eenum}{\end{enumerate}}	

    \newcommand{\lp}{\left(} 
    \newcommand{\rp}{\right)} 
    
    \newcommand{\D}{\mathrm{d}} 

\usepackage[textwidth=1.2cm]{todonotes}
\tikzset{inlinenotestyle/.append style={align=justify}}

\title[\dpg\ in field stars and in old-open clusters] {\emph{Kepler} red-clump stars in the field and in open clusters: constraints on core mixing}
\author[Bossini et al.]
{D. Bossini$^{1,2}$\thanks{E-mail: dbossini@bison.ph.bham.ac.uk},
A. Miglio$^{1,2}$,  M. Salaris$^{3}$,
{M. Vrard$^{4}$},                     
{S. Cassisi$^{5}$}, 
{B. Mosser$^{6}$},\newauthor
{J. Montalb\'an$^{7}$},  
{L. Girardi$^{8}$}, 
{A. Noels$^{9}$}, 
{A. Bressan$^{10}$}, 
{A. Pietrinferni$^{5}$}, {J. Tayar$^{11}$} 
\\
  $^{1}$ School of Physics and Astronomy, University of Birmingham, Edgbaston, Birmingham B15 2TT, United Kingdom \\
  $^{2}$ Stellar Astrophysics Centre, Department of Physics and Astronomy, Aarhus University, Aarhus, DK\\
  $^{3}$ Astrophysics Research Institute, Liverpool John Moores University, 146 Brownlow Hill, Liverpool L3 5RF, UK\\
  $^{4}$ Instituto de Astrof\'isica e Ci$\hat{e}$ncias do Espa\c{c}o, Universidade do Porto, CAUP, Rua das Estrelas, 4150-762 Porto,Portugal\\
  $^{5}$ Osservatorio Astronomico di Collurania -- INAF, via M. Maggini, 64100 Teramo, Italy\\
  $^{6}$ LESIA, Observatoire de Paris, PSL Research University, CNRS, Universit\'e Pierre et Marie Curie, Universit\'e Paris Diderot,\\  92195 Meudon, France\\
  $^{7}$ Dipartimento di Fisica e Astronomia, Universit\`a di Padova, Vicolo dell'Osservatorio 3, I-35122 Padova, Italy \\
  $^{8}$ Osservatorio Astronomico di Padova -- INAF, Vicolo dell'Osservatorio 5, I-35122 Padova, Italy \\
  $^{9}$ Institut d'Astrophysique et G\'eophysique de l'Universit\'e de Li\`ege, All\'ee du six Ao\^ut, 17 B-4000 Liege, Belgium\\
  $^{10}$ SISSA, via Bonomea 265, I-34136 Trieste, Italy\\
  $^{11}$ Department of Astronomy, Ohio State University, 140 W 18th Ave, OH 43210, USA\\   
 }
\date{Accepted 2017 May 3.  Received 2017 May 3; in original form  2017 April 7}

\pubyear{2017}

\begin{document}
\label{firstpage}

\pagerange{\pageref{firstpage}--\pageref{lastpage}}
\maketitle

\begin{abstract}
Convective mixing in Helium-core-burning (HeCB) stars is one of the outstanding issues in stellar modelling.
The precise asteroseismic measurements of gravity-modes period spacing (\dpg) has opened the door to detailed  studies of the near-core structure of such stars, which had not been possible before.
Here we provide stringent tests of various core-mixing scenarios against the largely unbiased population of red-clump stars belonging to the old open clusters monitored by {\it Kepler}, and by coupling the updated precise inference on \dpg\ in thousands field stars  with spectroscopic constraints.
We find that models with moderate overshooting successfully reproduce the range observed of \dpg\ in clusters.  In particular we show that there is no evidence for the need to extend the size of the adiabatically stratified core, at least at the beginning of the HeCB  phase. This conclusion is based primarily on ensemble studies of \dpg\ as a function of mass and metallicity.  
While \dpg\ shows no appreciable dependence on the mass, we have found a clear dependence of \dpg\ on metallicity, which is also supported by predictions from models. 

\end{abstract}



\vspace{-0.25cm}
\section{Introduction}
\label{sec:intro}
Modelling Helium-core-burning (HeCB) low-mass stars has proven to be complicated, given the lack of a detailed physical understanding of how energy and chemical elements are transported  in regions adjacent to convectively unstable cores. 
In particular, this phase is characterised by convective cores that tend to grow with evolution (hence generating sharp chemical profiles), and by the insurgence of a convectively unstable region separated from the core (called Helium-semiconvection, \citealt{Castellani_etal71b}). Overall, these uncertainties limit our ability to determine precise stellar properties (such as, e.g, mass and age), which are necessary in the context of studying stellar populations. Moreover, they generate uncertainties in model evolutionary tracks which affect a wide range of applications, including the theoretical calibration of red clump stars as distance indicators and the reliability of theoretical predictions about the following evolutionary stages such as the AGB and the WD ones \citep[see e.g.][]{Girardi16}.

It has been recently recognised that the gravity-mode period spacing (\dpg) measured in solar-like oscillating stars provides stringent constraints on core mixing processes in the Helium-core burning phase  \citep{Montalban_etal13, Bossini_etal15, Constantino_etal15}. 
In  our previous work \citep[][hereafter B15]{Bossini_etal15}, we investigated how key observational tracers of the near-core properties of HeCB stars (the luminosity of the AGB bump $L_{\rm AGBb}$ and, primarily, \dpg) depend on the core-mixing scheme adopted. By comparison with data from  \citet{APOKASC} and \citet{Mosser_etal12} we concluded, in agreement with  independent studies \citep{Constantino_etal15}, that no standard model can satisfactorily account for the period spacing of HeCB stars.  We then proposed a parametrised model (a moderate penetrative convection, i.e. $\sim0.5$ $H_p$ overshooting with adiabatic stratification in the extra-mixed region, see Sec. \ref{sec:models}) that is able to reproduce at the same time the observed distribution of \dpg\, and the $L_{\rm AGBb}$. 
However, we were prevented from drawing any further quantitative conclusions because  of the inherent limitation of comparing model predictions against a composite stellar population of less than $\sim 200$ stars, and  of potential biases affecting the measurement and detectability of the period spacing (as flagged, e.g. in  \citealt{Constantino_etal15}).

Here, we specifically address these concerns by studying \dpg\ of red-clump (RC) stars in old-open clusters, and by investigating the occurrence of any significant detection bias (Sec. \ref{sec:clusters}). 
Moreover, we take a further step and compare our predictions to the more stringent tests provided by  analysing the period-spacing structure \citep{Mosser_etal14,Vrard_etal16} coupled to spectroscopic constraints, which are now available for thousands of solar-like oscillating giants \citep{DR13}, which allows investigating trends of \dpg\, with mass and metallicity (Sec. \ref{sec:m&Z}).  

\vspace{-0.25cm}
\section{Models}
\label{sec:models}
We use the  stellar evolution code MESA  \citep{Paxton_etal13} to compute internal structures of stars during the helium-core-burning phase. The default set of physical inputs is described in \citet{Rodrigues_etal17}. We test several types of parametrised mixing schemes during the HeCB phase, which are classified based on the thermal stratification adopted in the region mixed beyond the \schw\ border, following the terminology introduced by \citet{Zahn91}. With the term {\it overshooting} (OV) we refer to  models in which the gradient of temperature in such region is radiative ($\gradT_\mathrm{,ovHe}=\gradr$) while {\it penetrative convection} (PC) indicates the cases where we assume an adiabatic gradient ($\gradT_\mathrm{,ovHe}=\grada$).
The size of the extra-mixed region is parametrised as $\aovhe\cdot\lambda$, where \aovhe\ is the overshooting parameter and $\lambda=\min{(H_{\rm p}, \rcc)}$ is the minimum between one pressure scale height, $H_{\rm p}$, and the radius of the convective core, \rcc.\footnote{This differs from the default parametrisation of overshooting in MESA, where $H_{\rm p}$ is instead considered equal to the minimum between \rcc\ and the mixing length $l_\mathrm{mlt}=\amlt H_{\rm p}$ (see \citealt{Deheuvels_etal16})} 
The mixing schemes tested in this work are:
\begin{itemize}
\item {\bf MOV}:  $\aovhe=0.5$, $\gradT_\mathrm{,ovHe}=\gradr$  (Moderate OV),
\item {\bf MPC}: $\aovhe=0.5$, $\gradT_\mathrm{,ovHe}=\grada$  (Moderate PC),
\item {\bf HOV}: $\aovhe=1.0$, $\gradT_\mathrm{,ovHe}=\gradr$ (High OV),
\item {\bf HPC}: $\aovhe=1.0$, $\gradT_\mathrm{,ovHe}=\grada$ (High PC).
\end{itemize}
In B15 we tested several of these schemes, and concluded that only MPC (computed using PARSEC, \citealt{Bressan_etal15}) was compatible with the observed \dpg\ and the luminosity distribution in the early AGB.
Regarding the large extra-mixing schemes ($\aovhe=1.0$) we found that only HOV had a good agreement with the observed \dpg. However, HOV fails to describe the luminosity distribution (too high $L_{\rm AGBb}$).
Finally, the plausibility of a bare-\schw\ scheme, here not included, which had already been ruled out theoretically \citep{Gabriel_etal14}, was also rejected by comparison with observations, including star counts in globular clusters \citep{Cassisi_etal03} and \dpg.
Compared to B15, we have modified the mixing scheme in models that develop  He-semiconvection zones (MOV, MPC).  We prevent the overshooting region to be (suddenly) attached to the He-semiconvection zone (which in MESA is treated as a convective region), by redefining \rcc\ to the minimum of \gradr. While this should be considered as an ad-hoc treatment with limited physical significance,  it provides a stable numerical scheme and mimics an efficient mixing in the He-semiconvective region. For further details see \citet{BossiniPhD}. 

\vspace{-0.25cm}
\section{Old-open clusters}
\label{sec:clusters}
Differently from field stars, open clusters are simple stellar populations, i.e coeval stars with the same initial chemical composition, and a similar mass for their evolved stars.
We can therefore perform a stringent test for the proposed mixing schemes in samples free of selection biases due to age, mass, and metallicity.

\begin{figure}
\centering
\resizebox{0.95\hsize}{!}{\includegraphics{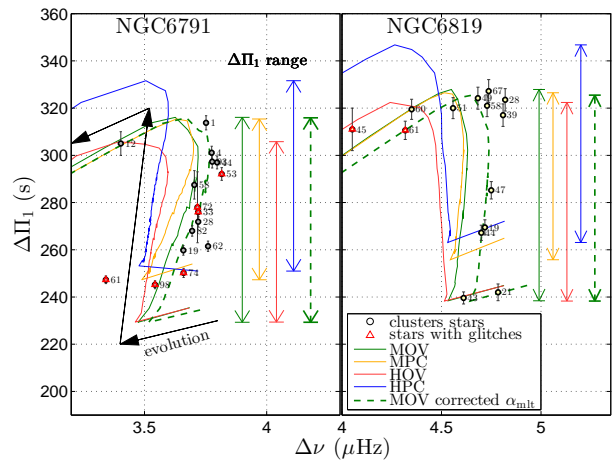}}
\vspace{-0.25cm}
\caption{Red-clump stars in NGC6791 and NGC6819 in a \dpg-\dnu\ diagram. Model predictions based on different near-core mixing schemes are shown by solid lines. The vertical lines indicate the range of \dpg\ covered by each mixing scheme. The dashed line shows MOV models computed with slightly ($\lesssim 3\%$) increased mixing-length parameters (compared to the solar-calibrated value). The red triangles mark the presence of buoyancy glitches which, to the first order, induce a modulation in \dpg\ with respect to the asymptotic value  (\citealt{Miglio_etal08}, \citealt{Cunha_etal15}, and Fig. 10 of \citealt{Mosser2015}),  potentially hampering an accurate inference of the asymptotic  \dpg.}
\label{fig:cluster_seismo}
\end{figure}

Our observational constraints on RC stars in the old-open clusters NGC6791 and NGC6819 are taken from \citet{Vrard_etal16} and, crucially, include measurements of the gravity-mode period spacing.
Among all the  stars observed by \kepler\ in NGC6791 and NGC6819, we exclude those not belonging to the RC ($\dpg<200$ s), stars  that are likely to be product of non-single evolution (over- / under-massive stars, Handberg \etal, submitted), and stars in which, according to \citet{Vrard_etal16}, the SNR is too low for a robust inference on \dpg\ (5 stars in NGC6819, and 3 in NGC6791). 
The final sample of RC stars in NGC6819 and NGC6791 consists of, respectively, 14 and 16 stars (see Table A.\ref{tab:NGC6819} and A.\ref{tab:NGC6791} for a complete list of targets). 
To compare data with theoretical predictions, we compute models representative of stars in the RC of the two clusters, adopting different extra-mixing schemes described in section \ref{sec:models}.
For NGC~6791 we calculate an evolutionary track with $M=1.15$ \Msun, $Z=0.0350$, and $Y=0.300$ \citep{Brogaard_etal11}, while for NGC~6819 we consider models with $M=1.60$ \Msun, $Z=0.0176$, and $Y=0.267$ (solar metallicity, Handberg \etal, submitted). 
Figure \ref{fig:cluster_seismo} shows the period spacings of the final samples of the observed RC stars as a function of their average large frequency separation (\dnu) and the comparison with our model predictions. 
\dnu\ in the models is computed from individual radial-mode frequencies (see \citealt{Rodrigues_etal17}).   
As in B15, the high-penetrative-convection scheme predicts a range of \dpg\ which is too high compared to the observations. The high-overshooting scheme, on the other hand, provides a range which is compatible with the observations, however,  it predicts too high a luminosity of the AGB bump. We note that models computed with the OV scheme have an upper limit of \dpg\ (during the HeCB phase) which does not monotonically increase with \aovhe\, but rather has a maximum at \aovhe$\simeq 0.6$. For higher values of \aovhe\ the He-semiconvection zone, that develops in the late phases of HeCB, remains separate from the convective core, allowing a larger radiative region,  thus effectively decreasing \dpg. 
This is the reason why in Fig. \ref{fig:cluster_seismo} MOV reaches higher \dpg\ values than HOV. 

The comparison between models and stars in NGC6791 and NGC6891 supports the conclusions reached in B15, i.e. that a moderate extra-mixed scheme reproduces well the maximum \dpg\ in the HeCB phase.
However, while in NGC6891 the MPC model cannot reach the small period spacings of two stars (21 and 43), that are likely to be early HeCB stars, the moderate overshooting scheme (MOV, green line in Fig. \ref{fig:cluster_seismo}) provides a better representation of the data. This model starts the HeCB with a lower \dpg, since the overshooting region is radiative, and reaches  \dpg\ as high as the MPC in the late HeCB, since the overall mixed core has $\gradT=\grada$ in both MOV and MPC schemes.
On the other hand, NGC6791 does not present early HeCB stars with small \dpg\ as predicted by the MOV scheme. 
A possible cause for this may be ascribed to the limited number of stars in the cluster, or to the three RC stars for which \dpg\ cannot be determined (see Table \ref{tab:NGC6791}).
The tentative evidence for such a discrepancy is supported by the general trend of the lower limit of \dpg\ with metallicity in field stars (see Sec. \ref{sec:m&Z} and Fig. \ref{fig:M_Z}). However, our sample around the cluster metallicity does not contain a sufficient number of stars to draw strong conclusions.

Offsets in \dnu\ between models and observations may be attributed to either small differences in the reference mass, or to systematic shifts in the effective temperature scale (due to e.g. uncertainties related to near-surface convection and to outer boundary condition), which modify the predicted photospheric radius, and hence \dnu\ ($\dnu\propto \sqrt{M/R^3}$). We notice that, for the models used in this study, e.g. increasing the solar-calibrated mixing-length parameter by $\lesssim 3\%$ is sufficient to recover a good agreement (see Fig. \ref{fig:cluster_seismo}). We stress that changing the outer boundary conditions / mixing length parameter has no impact on the predictions related to \dpg, which is determined by the near-core properties.

\vspace{-0.25cm}
\section{\dpg\ of Field stars}
\label{sec:m&Z}
\begin{figure}
\centering
\resizebox{0.95\hsize}{!}{\includegraphics{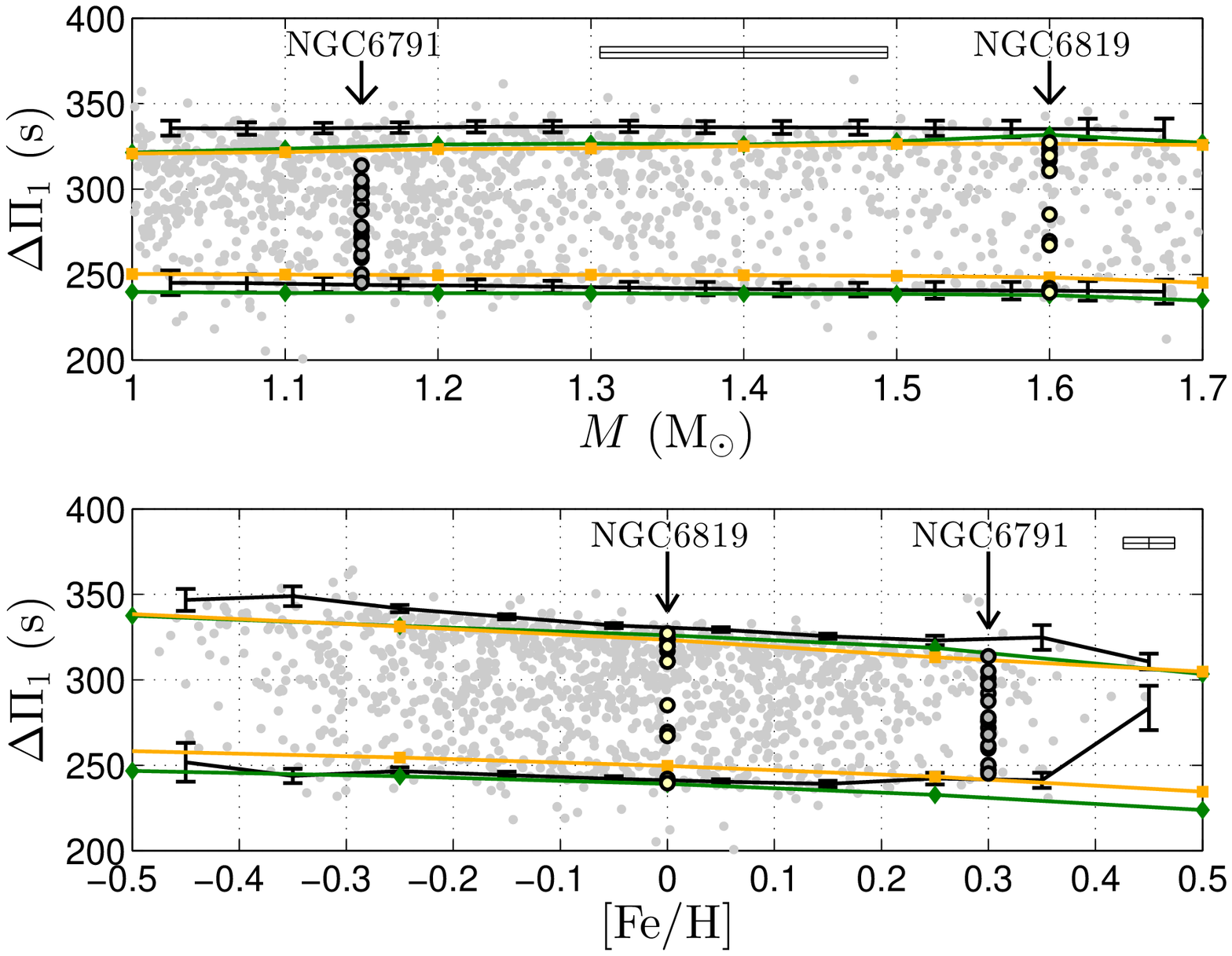}}
\vspace{-0.25cm}
\caption{Period spacing of HeCB stars with APOGEE DR13 spectroscopic parameters crossed with \citet{Vrard_etal16} plotted against mass (upper panel) and metallicity (lower panel). 
Black lines correspond to the 95th and 5th percentiles of the data distribution along \dpg, while green and orange lines represent the model predictions  (respectively MOV and MPC schemes) for \dpgmin\ and \dpgmax. 
An indication of the typical error on the data is visible in the top-right corner of each panel. 
NGC6791 (grey dots) and NGC6819 (yellow dots) cluster stars are also shown.}
\label{fig:M_Z}
\end{figure}

In this section we explore the effects of  mass and metallicity on the asymptotic period spacing of stars in the HeCB phase.
The dataset we use contains field stars with spectroscopic constrains available from APOGEE DR13 \citep{DR13} and \dpg\ reported in \citet{Vrard_etal16}. 
RC stars are selected looking for \dpg\ greater than $200$ s. 
The range of metallicity considered is $\feh\in[-0.5,0.5]$. 
We limit the mass range to \mbox{$M_\mathrm{seism}\in[1.0,1.7]$} \Msun\ in order to avoid stellar masses that are approaching  the secondary clump condition \citep[e.g., see][]{Girardi99}.
Figure \ref{fig:M_Z} shows the \dpg\ of the final selection plotted against the mass (upper panel) and metallicity (lower panel). 
It can be noticed that, in the interval considered, the period spacing is limited in a band between a maximum (\dpgmax) and a minimum (\dpgmin) value. 
To measure robustly the observed values of \dpgmax\ and \dpgmin\ we bin the dataset in mass and metallicity and for each bin we determine the 95th and 5th percentiles of the \dpg\ distribution (representing  \dpgmax\  and \dpgmin, respectively). 
In order to evaluate the uncertainties on the percentiles, taking in account also uncertainties on $M$,  \dpg\, and \feh,  we create 1000 realisations of the observed population. 
We use these to  calculate means and standard deviations of \dpgmax\ and \dpgmin\, which we then compare to model predictions (see the black lines in Fig. \ref{fig:M_Z}).

As evinced from the upper panel of Fig. \ref{fig:M_Z}, the  data show that the range of \dpg\ is largely independent of mass, while its upper and lower boundaries decrease with increasing metallicity.
To investigate whether models can account for such a behaviour, we compute a small grid of tracks that covers the range of mass and metallicity explored. 
In the lower panel of Fig.  \ref{fig:M_Z} we consider models at different metallicities with mass equal to 1.20 \Msun\ (close to the average mass of the $\sim 1.25$ \Msun\ of the observed distribution), while for the upper panel we fixed the metallicity to $\feh=0.00$ (mean observed value is $\feh=-0.034$) and vary the mass.  

We notice that models computed with the MOV  scheme are in overall good agreement with the observational constraints. \dpgmin, which is determined by the initial size of the adiabatically stratified core, is well reproduced by the MOV scheme, suggesting that models with PC are disfavoured at least in the initial phases of HeCB. \dpgmax, which depends on the core properties of the much more delicate advanced phases of HeCB, is also in good qualitative agreement with the MOV scheme. Interestingly, models also show decreasing \dpgmax\ and \dpgmin\ as metallicity increases.  The offset between the observed \dpgmax\ and the low-mass models (upper panel of Fig. \ref{fig:M_Z})  originates primarily from a metallicity effect.  Metallicity is not uniformly distributed across the range of masses considered, with small-mass stars being older and hence more metal poor, while the models shown in the upper panel are computed at fixed $Z$ (solar). 
\begin{figure}
\centering
\resizebox{.95\hsize}{!}{\includegraphics{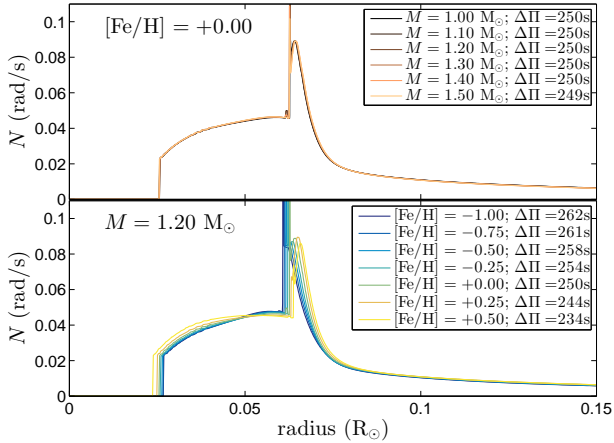}}
\vspace{-0.15cm}
\caption{\brunt\ frequency in the model grid presented in \citet{Rodrigues_etal17} at the start of the helium burning. $N$ profiles are shown on the upper panel for models with fixed metallicity ($\feh=0.00$), while in the lower panel for fixed a mass ($M=1.20$ \Msun).}
\label{fig:N}
\end{figure}

To interpret the behaviour of \dpgmax\ and \dpgmin\, it is worth recalling that the asymptotic period spacing of dipolar modes is related to the inverse of the integral of the \brunt\ frequency ($N$) over the radius (r) in the g-mode propagation cavity (\citealt{Tassoul80}):
\beq
    \dpg = \dfrac{2\pi^2}{\sqrt{2}}\lp\displaystyle\int^{r_2}_{r_1}\dfrac{N}{r}\D r\rp^{-1}.
    \label{eq:DPg}
\eeq
The region that mostly influences the integral in Eq. \ref{eq:DPg} is the radiative region near the centre (due to the dependency on $1/r$). Since $N$ is typically null in the deep fully convective regions, larger convective cores will lead to larger values of \dpg\ \citep{Montalban_etal13}. 
Looking at the \brunt\ frequency at the very beginning of the HeCB phase (Fig. \ref{fig:N}), we  notice that for a fixed metallicity all the profiles overlap, while visible differences are found by changing the metallicity. 
The reason behind this has to be searched in the mass of the helium rich core (\Mhe) at the beginning of HeCB, which determines the physical conditions of the central regions.
\Mhe\ is similar to the critical mass \Mhef, which is needed for the plasma to reach temperatures high enough to burn helium and start the helium flash.
For stars with masses in our range of interest, the critical mass \Mhef\ is the result of two competing mechanisms along the RGB: the central cooling due to the degeneracy and the hydrogen-burning shell that constantly deposits helium on the core. While the first is independent of  metallicy, the second has an efficiency that increases with $Z$.
The final effect tends to decrease \Mhef\ with increasing $Z$ \citep{CS13}.
Indeed, the different \Mhef\ and its properties influences the H-burning shell efficiency also in the HeCB phase, leading to high-$Z$ stars having a more efficient H-burning shell, which contributes more to the total luminosity than in the low-$Z$ stars. 
Therefore, in more metal rich stars the contribution of the He-core-burning to the whole energy budget is lower than in metal-poor ones. This occurrence has the consequence that metal-rich stars develop smaller convective core - and hence smaller \dpg\ - with respect low-Z stars.
On the other hand, low-mass stars with the same $Z$ end up with similar \Mhef\ and therefore similar helium core during the HeCB and similar \dpg. However, this is true only for $M\lesssim1.5$-$1.7$ \Msun\ (depending on $Z$). Above this value, \Mhef\ starts to decrease since we approach the secondary clump (the degeneracy of the He core decreases and hence \Mhef, \citealt{Girardi99}). 

\begin{figure}
\centering
\resizebox{0.95\hsize}{!}{\includegraphics{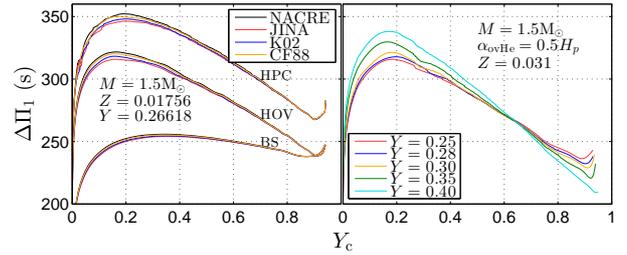}}
\vspace{-0.25cm}
\caption{{\it Left panel:} RC-\dpg\ evolution as a function of central helium mass of a $M=1.50$ \Msun\ solar metallicity star with three different mixing schemes and four different \coreaction\ reaction rates.  {\it Right panel:} RC-\dpg\ evolution as a function of central helium mass fraction of a $M=1.50$ \Msun\ star computed assuming three different combinations of $Z$ and $Y$.}
\label{fig:CO_Y}
\end{figure}

An additional test is made to quantify the effect of the initial helium on \dpg. In our grid $Y$ is in fact coupled with $Z$ via linear chemical enrichment relation \citep[see][]{Rodrigues_etal17}.
To decouple the effect of $Z$ and $Y$ we compute five tracks of mass $1.50\, \Msun$ and $Z=0.031$ ($\feh=0.25$) but with initial helium $Y=0.25,0.28,0.30,0.35,0.40$ (with $Y=0.28$ as our default value).  
Figure \ref{fig:CO_Y}, right panel, shows the evolution of \dpg\ with central helium for the 5 tracks. 
We notice that the effect on \dpgmax\ and \dpgmin\ grows linearly with $Y$. However for variation of $\Delta Y=\sim0.02$ from the default value  ($Y=0.25$ and $Y=0.30$) the deviation on \dpgmax\ (\dpgmin)  lays within the observed uncertainty on \dpg\ ($\sim3$-$4$ s). The deviation we tested is compatible with typical spread on disk populations \citep{Casagrande_etal07}.  
Nevertheless, the effect becomes substantial for extreme enrichments, e.g. in bulge or globular clusters where populations with very different He abundances and same metallicity may coexist \citep{Renzini_etal15}.   

\vspace{-0.25cm}
\section{Additional uncertainties on \dpg: \coreaction\ nuclear reaction rate}\label{sec:co_DP}
As shown above  \dpg\ is strongly dependent on assumptions related to core convection and metallicity, however, additional parameters and uncertainties may also impact on \dpg, in particular the \coreaction\ reaction rate, which, along with \trialfa, plays a fundamental role, especially  at the end of the HeCB \citep[e.g. see][]{Straniero_etal03, Cassisi_etal03, Constantino_etal15}.

We compute a series of HeCB evolutionary tracks ($M=1.5$ \Msun, solar abundance) in which we adopt four \coreaction\ reaction rates in conjunction with 3 mixing schemes: Bare-\schw\ model (BS), $1.0$ $H_p$ step function overshooting (HOV), and 1 $H_p$ penetrative convection (HPC).
The \coreaction\ reaction rates considered are the tabulated values given by JINA \citep{JINA}, K02 \citep{Kunz_etal02}, CF88 \citep{CF88}, and NACRE \citep{NACRE} and already made available in MESA. 
While no difference can be noticed at the beginning of the phase, the impact of the different mixing schemes is evident at the maximum period spacing (end of the HeCB), where HOV tracks show a scatter of around $6$-$7$ s between them, compared to only $\sim2$ s for BS (Figure \ref{fig:CO_Y}).  
We therefore expect an uncertainty  between 6 and 2 s on the MPC and MOV models. This value is comparable with  the observed average  \dpg\ uncertainty for clump stars ($\sim4$ s).

\vspace{-0.25cm}
\section{Discussion and Conclusion}
\label{sec:conclusion}
The precise measurements of gravity-modes period spacing (\dpg) in thousands of He-core-burning stars has opened the door to detailed  studies of the near core structure of such stars, which had not been possible before \citep{Montalban_etal13, Bossini_etal15, Constantino_etal15}

Here we provide additional stringent tests of the mixing schemes by stress-testing the models presented in B15 against results on the simple population of RC stars belonging to the old open clusters monitored by {\it Kepler}, and making use of the updated precise inference on \dpg\ presented by \citet{Vrard_etal16} coupled with spectroscopic constraints from APOGEE DR13 \citep{DR13}.

We find that in clusters \dpg\ is measured in all RC stars with few exceptions (as discussed in Section \ref{sec:clusters}, Table \ref{tab:NGC6819} and \ref{tab:NGC6791}), and that models with moderate overshooting can reproduce the range of period spacing observed.  
In particular our models do not support the need to extend the size of the adiabatically stratified core, at least at the beginning of the HeCB  phase. 
This conclusion is based primarily on ensemble studies of \dpg\ as a function of mass and metallicity, where we could also show that models successfully reproduce the main trends (or their absence). 
While \dpg\ shows no appreciable dependence on the mass, we have found a clear dependence of \dpg\ on metallicity (Figure \ref{fig:M_Z}) also shown by the models, which strengthens even further the result on the clusters.  
We complement the study  by considering how theoretically predicted \dpg\ depends on the  initial helium mass fraction and on the nuclear cross sections adopted in the models, and conclude that the adopted mixing scheme and metallicity are the dominant effects.  

The parametrised model presented here should  be considered as a first approximation which broadly reproduces the inferred average asymptotic period spacing. Significant improvements can be made by looking at signatures of sharp-structure variations \citep{Mosser2015, Cunha_etal15, Bossini_etal15} which will enable to test in greater detail the chemical and temperature stratification near the edge of the convective core, providing additional indications that will eventually be compared with more realistic and physically justified models of convection in this key stellar evolutionary phases.

\vspace{-0.4cm}
\section*{Acknowledgments}
AM acknowledges the support of the UK Science and Technology Facilities Council (STFC). We acknowledge the support from the PRIN INAF 2014 -- CRA 1.05.01.94.05.  
JM  acknowledges support from the ERC Consolidator Grant funding scheme ({\em project STARKEY}, G.A. n. 615604). 
MV acknowledges funding by the Portuguese Science foundation through the grant with reference CIAAUP 03/2016BPD, in the context of the project FIS/04434, cofunded by FEDER through the program COMPLETE.
SC acknowledges financial support from PRIN-INAF2014 (PI: S. Cassisi)
\vspace{-0.3cm}
\bibliographystyle{mnras}
\bibliography{Bossini_etal} 
\newpage
\onecolumn
\begin{table}
    \centering

    \caption{\label{tab:NGC6819} NGC6819 stellar catalogue. Stellar classes were chosen considering the classification in \citet{Stello_etal11} and \dpg\ in \citet{Vrard_etal16}.f}
    \begin{tiny}               
        \begin{tabular}{cccccccccl}
\hline
             N &            KIC &         \numax &           \dnu &           \dpg & $\sigma_{\Delta\Pi_1}$ &   Vrard et al. &          class &       selected &          notes \\
               &                &      ($\mu$Hz) &      ($\mu$Hz) &            (s) &                    (s) &         (2016) &                &       RC stars &                \\
\hline
             1 &        4937011 &            --- &            --- &            --- &                    --- &            --- &            RGB &            --- &                \\
             2 &        4937056 &          46.10 &           4.76 &            --- &                    --- &            yes &        unclear &            --- & low l=1 structure \\
             3 &        4937257 &            --- &            --- &            --- &                    --- &            --- &            RGB &            --- &                \\
             4 &        4937576 &          32.95 &           3.56 &            --- &                    --- &            yes &            RGB &            --- &                \\
             5 &        4937770 &          94.30 &           7.83 &         160.70 &                   1.99 &            yes &        unclear &            --- & Possible second clump.  \\
             6 &        4937775 &          89.92 &           7.33 &         226.50 &                   4.61 &            yes &        unclear &            --- & little l=1 structure. No photometric clump   \\
             7 &        5023732 &          27.08 &           3.12 &            --- &                    --- &            yes &            RGB &            --- &                \\
             8 &        5023845 &         109.94 &           8.96 &            --- &                    --- &            yes &            RGB &            --- &                \\
             9 &        5023889 &            --- &            --- &            --- &                    --- &            --- &            RGB &            --- &                \\
            10 &        5023931 &          50.57 &           4.92 &            --- &                    --- &            yes &            RGB &            --- & little l=1 structure  \\
            11 &        5023953 &          48.75 &           4.74 &         299.00 &                   4.36 &            yes &             RC &             no & $M_\mathrm{seismo}=1.83\Msun$ \\
            12 &        5024043 &          55.98 &           5.64 &          61.00 &                   2.00 &            yes &            RGB &            --- &                \\
            13 &        5024143 &         122.84 &           9.68 &          68.40 &                   6.59 &            yes &            RGB &            --- &                \\
            14 &        5024240 &         153.85 &          12.00 &            --- &                    --- &            yes &            RGB &            --- &                \\
            15 &        5024268 &            --- &            --- &            --- &                    --- &            --- &            RGB &            --- &                \\
            16 &        5024272 &            --- &            --- &            --- &                    --- &            --- &            RGB &            --- &                \\
            17 &        5024297 &          46.24 &           4.60 &            --- &                    --- &            yes &            RGB &            --- &                \\
            18 &        5024312 &          96.68 &           8.13 &            --- &                    --- &            yes &            RGB &            --- &                \\
            19 &        5024327 &          44.18 &           4.72 &         269.50 &                   3.21 &            yes &             RC &            yes &                \\
            20 &        5024329 &            --- &            --- &            --- &                    --- &            --- &            RGB &            --- &                \\
            21 &        5024404 &          47.09 &           4.78 &         242.00 &                   3.54 &            yes &             RC &            yes &                \\
            22 &        5024405 &          98.89 &           8.29 &            --- &                    --- &            yes &            RGB &            --- &                \\
            23 &        5024414 &          78.17 &           6.46 &         280.70 &                   6.16 &            yes &             RC &             no & $M_\mathrm{seismo}=2.63\Msun$ \\
            24 &        5024456 &           3.86 &           0.70 &            --- &                    --- &            yes &            RGB &            --- &                \\
            25 &        5024476 &          65.94 &           5.74 &         298.00 &                   3.79 &            yes &             RC &             no & $M_\mathrm{seismo}=2.38\Msun$ \\
            26 &        5024512 &          72.97 &           6.70 &            --- &                    --- &            yes &            RGB &            --- &                \\
            27 &        5024517 &          50.13 &           4.94 &         319.20 &                   5.11 &            yes &            RGB &            --- & non-photometric member \\
            28 &        5024582 &          46.30 &           4.82 &         323.50 &                   4.76 &            yes &             RC &            yes &                \\
            29 &        5024583 &          37.89 &           3.91 &            --- &                    --- &            yes &            RGB &            --- &                \\
            30 &        5024601 &          32.30 &           3.68 &            --- &                    --- &            yes &             RC &            --- &  Very low l=1  \\
            31 &        5024750 &          12.74 &           1.80 &            --- &                    --- &            yes &            RGB &            --- &                \\
            32 &        5024851 &           4.06 &           0.75 &            --- &                    --- &            yes &            RGB &            --- &                \\
            33 &        5024870 &            --- &            --- &            --- &                    --- &            --- &            RGB &            --- &                \\
            34 &        5024967 &          44.97 &           4.71 &            --- &                    --- &            yes &             RC &            --- &  Very low l=1  \\
            35 &        5024984 &            --- &            --- &            --- &                    --- &            --- &            RGB &            --- &                \\
            36 &        5111718 &         135.49 &          10.59 &          87.53 &                   1.06 &            yes &            RGB &            --- &                \\
            37 &        5111820 &            --- &            --- &            --- &                    --- &            --- &            RGB &            --- &                \\
            38 &        5111940 &          52.79 &           5.20 &            --- &                    --- &            yes &            RGB &            --- &                \\
            39 &        5111949 &          47.35 &           4.81 &         317.00 &                   4.76 &            yes &             RC &            yes &                \\
            40 &        5112072 &         125.27 &          10.08 &          92.40 &                   0.91 &            yes &            RGB &            --- &                \\
            41 &        5112288 &          46.94 &           4.77 &            --- &                    --- &            yes &             RC &             no &  Very low l=1  \\
            42 &        5112361 &          67.63 &           6.19 &          91.40 &                   0.90 &            yes &            RGB &            --- &                \\
            43 &        5112373 &          43.84 &           4.61 &         239.60 &                   2.57 &            yes &             RC &            yes &                \\
            44 &        5112387 &          44.91 &           4.70 &         267.20 &                   3.21 &            yes &             RC &            yes &                \\
            45 &        5112401 &          36.50 &           4.05 &         311.00 &                   9.00 &            yes &             RC &            yes & presence of glitch  \\
            46 &        5112403 &         141.73 &          11.18 &          86.80 &                   1.07 &            yes &            RGB &            --- &                \\
            47 &        5112467 &          45.15 &           4.75 &         285.20 &                   3.67 &            yes &             RC &            yes &                \\
            48 &        5112481 &           5.18 &           0.92 &            --- &                    --- &            yes &            RGB &            --- &                \\
            49 &        5112491 &          44.25 &           4.68 &         324.20 &                   4.65 &            yes &             RC &            yes &                \\
            50 &        5112558 &            --- &            --- &            --- &                    --- &            --- &            RGB &            --- &                \\
            51 &        5112730 &          43.60 &           4.56 &         320.00 &                   4.46 &            yes &             RC &            yes &                \\
            52 &        5112734 &          40.65 &           4.16 &            --- &                    --- &            yes &            RGB &            --- &                \\
            53 &        5112741 &            --- &            --- &            --- &                    --- &            --- &            RGB &            --- &                \\
            54 &        5112744 &          43.97 &           4.44 &            --- &                    --- &            yes &            RGB &            --- &                \\
            55 &        5112751 &           1.32 &           0.39 &            --- &                    --- &            yes &             RC &             no &  Very low l=1  \\
            56 &        5112786 &           7.70 &           1.17 &            --- &                    --- &            yes &            RGB &            --- &                \\
            57 &        5112880 &          25.43 &           2.82 &            --- &                    --- &            yes &            RGB &            --- &                \\
            58 &        5112938 &          44.54 &           4.73 &         321.00 &                   4.59 &            yes &             RC &            yes &                \\
            59 &        5112948 &          42.28 &           4.31 &            --- &                    --- &            yes &            RGB &            --- &                \\
            60 &        5112950 &          41.59 &           4.35 &         319.50 &                   4.26 &            yes &             RC &            yes &                \\
            61 &        5112974 &          40.08 &           4.32 &         310.60 &                   3.87 &            yes &             RC &            yes & presence of glitch \\
            62 &        5113041 &          37.13 &           4.01 &            --- &                    --- &            yes &            RGB &            --- &                \\
            63 &        5113061 &           4.53 &           0.84 &            --- &                    --- &            yes &            RGB &            --- &                \\
            64 &        5113441 &         154.68 &          11.76 &          89.65 &                   1.24 &            yes &            RGB &            --- &                \\
            65 &        5199859 &           0.70 &           0.15 &            --- &                    --- &            yes &            RGB &            --- &                \\
            66 &        5200088 &            --- &            --- &            --- &                    --- &            --- &            RGB &            --- &                \\
            67 &        5200152 &          45.71 &           4.74 &         327.20 &                   4.89 &            yes &             RC &            yes &                \\
\hline
\end{tabular}

    \end{tiny}
\end{table}
\begin{table}
    \centering
    \caption{\label{tab:NGC6791} NGC6791 stellar catalogue. Stellar classes were chosen considering the classification in \citet{Stello_etal11} and \dpg\ in \citet{Vrard_etal16}. }
    \begin{tiny}               
        \begin{tabular}{cccccccccl}
\hline
             N &            KIC &         \numax &           \dnu &           \dpg & $\sigma_{\Delta\Pi_1}$ &   Vrard et al. &          class &       selected &          notes \\
               &                &      ($\mu$Hz) &      ($\mu$Hz) &            (s) &                    (s) &         (2016) &                &       RC stars &                \\
\hline
             1 &        2297384 &          30.49 &           3.75 &         313.78 &                   3.00 &            yes &             RC &            yes &                \\
             2 &        2297574 &            --- &            --- &            --- &                    --- &            --- &            RGB &            --- &                \\
             3 &        2297793 &            --- &            --- &            --- &                    --- &            --- &            RGB &            --- &                \\
             4 &        2297825 &          30.43 &           3.77 &         301.10 &                   2.76 &            yes &             RC &            yes &                \\
             5 &        2298097 &            --- &            --- &            --- &                    --- &            --- &            RGB &            --- &                \\
             6 &        2435987 &          38.07 &           4.22 &            --- &                    --- &            yes &            RGB &            --- &                \\
             7 &        2436097 &          42.06 &           4.54 &            --- &                    --- &            yes &            RGB &            --- &                \\
             8 &        2436209 &          57.01 &           5.76 &          67.30 &                   2.46 &            yes &            RGB &            --- &                \\
             9 &        2436291 &            --- &            --- &            --- &                    --- &            --- &            RGB &            --- &                \\
            10 &        2436332 &          28.29 &           3.40 &            --- &                    --- &            yes &            RGB &            --- &                \\
            11 &        2436376 &            --- &            --- &            --- &                    --- &            --- &            RGB &            --- &                \\
            12 &        2436417 &          27.07 &           3.40 &         305.00 &                   5.00 &            yes &             RC &            yes &                \\
            13 &        2436458 &          37.08 &           4.17 &            --- &                    --- &            yes &            RGB &            --- &                \\
            14 &        2436540 &          57.76 &           5.83 &            --- &                    --- &            yes &            RGB &            --- &                \\
            15 &        2436593 &          24.96 &           3.56 &            --- &                    --- &            yes &            RGB &            --- &    binary star \\
            16 &        2436676 &         131.86 &          11.35 &          80.60 &                   1.16 &            yes &            RGB &            --- &                \\
            17 &        2436688 &          76.01 &           7.28 &            --- &                    --- &            yes &            RGB &            --- &                \\
            18 &        2436715 &            --- &            --- &            --- &                    --- &            --- &            RGB &            --- &                \\
            19 &        2436732 &          30.27 &           3.66 &         259.76 &                   2.04 &            yes &             RC &            yes &                \\
            20 &        2436759 &          32.63 &           3.73 &            --- &                    --- &            yes &            RGB &            --- &                \\
            21 &        2436804 &            --- &            --- &            --- &                    --- &            --- &            RGB &            --- &                \\
            22 &        2436814 &          24.51 &           3.13 &            --- &                    --- &            yes &            RGB &            --- &                \\
            23 &        2436818 &          97.32 &           8.84 &          76.10 &                   0.56 &            yes &            RGB &            --- &                \\
            24 &        2436824 &          34.03 &           3.87 &            --- &                    --- &            yes &            RGB &            --- &                \\
            25 &        2436842 &            --- &            --- &            --- &                    --- &            --- &            RGB &            --- &                \\
            26 &        2436900 &          35.62 &           4.07 &            --- &                    --- &            yes &            RGB &            --- &                \\
            27 &        2436912 &          29.79 &           3.73 &            --- &                    --- &            yes &             RC &             no & suppressed l=1  \\
            28 &        2436944 &          30.86 &           3.72 &         271.90 &                   2.28 &            yes &             RC &            yes &                \\
            29 &        2436954 &          34.48 &           4.16 &            --- &                    --- &            yes &            RGB &            --- &                \\
            30 &        2436981 &            --- &            --- &            --- &                    --- &            --- &            RGB &            --- &                \\
            31 &        2437033 &            --- &            --- &            --- &                    --- &            --- &            RGB &            --- &                \\
            32 &        2437040 &          25.49 &           3.08 &            --- &                    --- &            yes &            RGB &            --- &                \\
            33 &        2437103 &          28.46 &           3.72 &         276.00 &                   2.17 &            yes &             RC &            yes & presence of glitch \\
            34 &        2437112 &            --- &            --- &            --- &                    --- &            --- &            RGB &            --- &                \\
            35 &        2437171 &            --- &            --- &            --- &                    --- &            --- &            RGB &            --- &                \\
            36 &        2437178 &            --- &            --- &            --- &                    --- &            --- &            RGB &            --- &                \\
            37 &        2437209 &            --- &            --- &            --- &                    --- &            --- &            RGB &            --- &                \\
            38 &        2437240 &          45.56 &           4.86 &          63.60 &                   1.68 &            yes &            RGB &            --- &                \\
            39 &        2437261 &            --- &            --- &            --- &                    --- &            --- &            RGB &            --- &                \\
            40 &        2437270 &          69.36 &           6.54 &          62.60 &                   2.75 &            yes &            RGB &            --- &                \\
            41 &        2437296 &            --- &            --- &            --- &                    --- &            --- &            RGB &            --- &                \\
            42 &        2437325 &          93.60 &           8.54 &          75.30 &                   0.53 &            yes &            RGB &            --- &                \\
            43 &        2437340 &           8.44 &           1.30 &            --- &                    --- &            yes &            RGB &            --- &                \\
            44 &        2437353 &          31.25 &           3.80 &         297.00 &                   2.73 &            yes &             RC &            yes &                \\
            45 &        2437394 &         159.70 &          12.99 &            --- &                    --- &            yes &            RGB &            --- &                \\
            46 &        2437402 &          46.41 &           4.84 &            --- &                    --- &            yes &            RGB &            --- &                \\
            47 &        2437413 &            --- &            --- &            --- &                    --- &            --- &            RGB &            --- &                \\
            48 &        2437443 &            --- &            --- &            --- &                    --- &            --- &            RGB &            --- &                \\
            49 &        2437444 &          18.83 &           2.48 &            --- &                    --- &            yes &            RGB &            --- &                \\
            50 &        2437488 &          64.77 &           6.30 &            --- &                    --- &            yes &            RGB &            --- &                \\
            51 &        2437496 &           4.43 &           0.86 &            --- &                    --- &            yes &            RGB &            --- &                \\
            52 &        2437507 &          20.41 &           2.62 &            --- &                    --- &            yes &            RGB &            --- &                \\
            53 &        2437564 &          32.02 &           3.82 &         292.10 &                   2.69 &            yes &             RC &            yes & presence of glitch \\
            54 &        2437589 &          46.11 &           4.63 &            --- &                    --- &            yes &        unclear &            --- & Uncertain \dpg. probably RGB. \\
            55 &        2437624 &            --- &            --- &            --- &                    --- &            --- &            RGB &            --- &                \\
            56 &        2437630 &            --- &            --- &            --- &                    --- &            --- &            RGB &            --- &                \\
            57 &        2437653 &          74.58 &           7.07 &            --- &                    --- &            yes &            RGB &            --- &                \\
            58 &        2437698 &          29.78 &           3.70 &         287.50 &                   6.01 &            yes &             RC &            yes &                \\
            59 &        2437781 &          85.46 &           7.88 &            --- &                    --- &            yes &            RGB &            --- &                \\
            60 &        2437792 &            --- &            --- &            --- &                    --- &            --- &            RGB &            --- &                \\
            61 &        2437804 &          26.71 &           3.34 &         247.30 &                   1.63 &            yes &             RC &            yes &                \\
            62 &        2437805 &          31.93 &           3.76 &         261.50 &                   2.16 &            yes &             RC &            yes &                \\
            63 &        2437816 &          17.78 &           2.37 &            --- &                    --- &            yes &            RGB &            --- &                \\
            64 &        2437851 &          12.37 &           1.90 &            --- &                    --- &            yes &            RGB &            --- &                \\
            65 &        2437897 &            --- &            --- &            --- &                    --- &            --- &            RGB &            --- &                \\
            66 &        2437930 &            --- &            --- &            --- &                    --- &            --- &            RGB &            --- &                \\
            67 &        2437933 &         107.46 &           9.45 &          76.80 &                   0.63 &            yes &            RGB &            --- &                \\
            68 &        2437957 &          93.15 &           8.57 &            --- &                    --- &            yes &            RGB &            --- &                \\
            69 &        2437965 &           7.20 &           1.28 &            --- &                    --- &            yes &            RGB &            --- &                \\
            70 &        2437972 &          85.43 &           7.89 &          69.10 &                   0.41 &            yes &            RGB &            --- &                \\
            71 &        2437976 &          89.62 &           8.21 &          75.50 &                   0.51 &            yes &            RGB &            --- &                \\
            72 &        2437987 &          29.95 &           3.72 &         278.00 &                  15.00 &            yes &             RC &            yes & presence of glitch \\
            73 &        2438038 &          62.25 &           6.18 &          66.40 &                   2.65 &            yes &            RGB &            --- &                \\
            74 &        2438051 &          30.46 &           3.66 &         250.20 &                   1.93 &            yes &             RC &            yes & presence of glitch \\
            75 &        2438053 &            --- &            --- &            --- &                    --- &            --- &            RGB &            --- &                \\
            76 &        2438140 &          70.96 &           6.79 &          67.30 &                   3.20 &            yes &            RGB &            --- &                \\
            77 &        2438192 &            --- &            --- &            --- &                    --- &            --- &            RGB &            --- &                \\
            78 &        2438333 &          61.09 &           6.11 &          65.00 &                   2.48 &            yes &            RGB &            --- &                \\
            79 &        2438421 &           0.67 &           0.21 &            --- &                    --- &            yes &            RGB &            --- &                \\
            80 &        2568519 &            --- &            --- &            --- &                    --- &            --- &            RGB &            --- &                \\
            81 &        2568916 &           0.45 &           0.23 &            --- &                    --- &            yes &             RC &             no &  very poor SNR \\
            82 &        2569055 &          30.49 &           3.69 &         268.00 &                   2.22 &            yes &             RC &            yes &                \\
            83 &        2569126 &            --- &            --- &            --- &                    --- &            --- &            RGB &            --- &                \\
            84 &        2569360 &          21.32 &           2.76 &            --- &                    --- &            yes &            RGB &            --- &                \\
            85 &        2569488 &           0.54 &           0.23 &            --- &                    --- &            yes &             RC &             no &  very poor SNR \\
            86 &        2569618 &          56.41 &           5.70 &            --- &                    --- &            yes &            RGB &            --- &                \\
            87 &        2569650 &            --- &            --- &            --- &                    --- &            --- &            RGB &            --- &                \\
            88 &        2569673 &            --- &            --- &            --- &                    --- &            --- &            RGB &            --- &                \\
            89 &        2569712 &            --- &            --- &            --- &                    --- &            --- &            RGB &            --- &                \\
            90 &        2569752 &            --- &            --- &            --- &                    --- &            --- &            RGB &            --- &                \\
            91 &        2569912 &            --- &            --- &            --- &                    --- &            --- &            RGB &            --- &                \\
            92 &        2569935 &           5.20 &           0.98 &            --- &                    --- &            yes &            RGB &            --- &                \\
            93 &        2569945 &          30.13 &           3.78 &         297.30 &                   2.66 &            yes &             RC &            yes &                \\
            94 &        2570094 &          67.83 &           6.55 &          71.05 &                   0.34 &            yes &            RGB &            --- &                \\
            95 &        2570134 &            --- &            --- &            --- &                    --- &            --- &            RGB &            --- &                \\
            96 &        2570144 &            --- &            --- &            --- &                    --- &            --- &            RGB &            --- &                \\
            97 &        2570172 &          75.15 &           7.09 &            --- &                    --- &            yes &            RGB &            --- &                \\
            98 &        2570214 &          28.05 &           3.54 &         245.20 &                   1.69 &            yes &             RC &            yes & presence of glitch \\
            99 &        2570244 &         105.70 &           9.28 &          76.85 &                   0.62 &            yes &            RGB &            --- &                \\
           100 &        2570263 &            --- &            --- &            --- &                    --- &            --- &            RGB &            --- &                \\
           101 &        2570384 &          47.54 &           4.84 &            --- &                    --- &            yes &            RGB &            --- &                \\
           102 &        2570518 &          46.45 &           4.98 &            --- &                    --- &            yes &            RGB &            --- &                \\
           103 &        2570519 &            --- &            --- &            --- &                    --- &            --- &            RGB &            --- &                \\
           104 &        2579142 &            --- &            --- &            --- &                    --- &            --- &            RGB &            --- &                \\
           105 &        2582664 &            --- &            --- &            --- &                    --- &            --- &            RGB &            --- &                \\
           106 &        2585397 &            --- &            --- &            --- &                    --- &            --- &            RGB &            --- &                \\
\hline
\end{tabular}

    \end{tiny}
\end{table}

\bsp	
\label{lastpage}
\end{document}